\documentstyle[12pt]{article}

\newcommand{\Dslash}{\mbox{ $\not \!\! D$}}

\newcommand{\beg}{\begin{equation}}
\newcommand{\ene}{\end{equation}}

\makeatletter
\@addtoreset{equation}{section}
\makeatother

\relax
\def\np#1#2#3 {Nucl. Phys. \underline{#1} (19#2) #3}
\def\prl#1#2#3 {Phys. Rev. Lett. \underline{#1} (19#2) #3}
\def\prev#1#2#3 {Phys. Rev. \underline{#1} (19#2) #3}
\def\pl#1#2#3 {Phys. Lett. \underline{#1} (19#2) #3}
\def\rmp#1#2#3 {Rev. Mod. Phys. \underline{#1} (19#2) #3}
\def\prep#1#2#3 {Phys. Rep. \underline{#1} (19#2) #3}
\def\spu#1#2#3 {Sov. Phys.-Usp. \underline{#1} (19#2) #3}
\def\sjnp#1#2#3 {Sov. J. Nucl. Phys. \underline{#1} (19#2) #3}
\def\jetp#1#2#3 {JETP Lett. \underline{#1} (19#2) #3}
\def\apj#1#2#3 {Astrophys. J. \underline{#1} (19#2) #3}
\def\apjl#1#2#3 {Astrophys. J. Lett. \underline{#1} (19#2) #3}
\def\ib#1#2#3 {{\it ibid.} \underline{#1} (19#2) #3}
\def\nat#1#2#3 {Nature (London) \underline{#1} (19#2) #3}
\def\ap#1#2#3 {Ann. Phys. (NY) \underline{#1} (19#2) #3}
\def\nc#1#2#3 {Nuovo Cim. \underline{#1} (19#2) #3}
\def\zp#1#2#3 {Zeit. Phys. \underline{#1} (19#2) #3}
\def\ar#1#2#3 {Ann. Rev. Nucl. Part. Sci. \underline{#1} (19#2) #3}
\def\prs#1#2#3 {Proc. Roy. Soc. \underline{#1} (19#2) #3}
\def\pcps#1#2#3 {Proc. Cam. Phil. Soc. \underline{#1} (#2) #3}
\def\rpp#1#2#3 {Rep. Prog. Phys. \underline{#1} (19#2) #3}
\def\cpc#1#2#3 {Computer Phys. Comm. \underline{#1} (19#2) #3}
\def\jchp#1#2#3 {J. Chem. Phys. \underline{#1} (19#2) #3}
\def\ptp#1#2#3 {Prog. Th. Phys. \underline{#1} (19#2) #3}
\def\beq{\begin{equation}}
\def\eeq{\end{equation}}

\relax
\begin{document}

\title{Full QCD on APE100 Machines.}

\author{
Stefano ANTONELLI,  Marco BELLACCI, \\
Andrea DONINI, Renata SARNO  \\
Dipartimento di Fisica, Universit\`a di Roma "La Sapienza",\\
P.~A.~Moro, 00185 Roma, Italy\\
and \\
INFN, Sezione di Roma \\
}
\maketitle

\begin{abstract}
We present the first tests and results from a study of QCD with two
flavours of dynamical Wilson fermions using the Hybrid Monte Carlo
Algorithm (HMCA) on APE100 machines. \par
Initially we have tested the algorithm without fermions to produce
SU(3) pure gauge configurations. The simulations have been performed
on a $6^4$ lattice at $\beta=5.7$ and $\beta=6.0$. \par
Then we started with the full QCD simulations.
The HMCA parameters have been tuned using a wide class of tests
performed on a $8^4$ and on $12^3\times 32$ lattices. First results for the
mesons correlations functions on $12^3 \times 32$ lattice are obtained at
$\beta=5.3$. \par
In this paper we are briefly facing some technical aspects: we discuss
about the inversion algorithm for the fermionic operator, we underline
the methods used to overcome the problems arising using a 32 bits
machine and we discuss the implementation of a new random number generator
for APE100 machines. \par
Finally we propose different scenarios for the simulation of physical
observables, with respect to the memory capacity and speed of different
APE100 configurations.
\end{abstract}
ROME Prep. 93/972 \\
\vfill
\newpage

\section{ Introduction.}
Numerical simulations of Quantum Field Theories on the lattice by
means of Monte Carlo  methods have become an important tool to
understand and predict non-perturbative phenomena in particle physics.
The quantitative study of some aspects of quantum field theories,
for example the determination of the hadronic mass spectrum of QCD,
requires non-perturbative approach as such as  the lattice one. \par

Up to now most of the numerical simulations in lattice QCD have been
performed in the quenched approximation and results on large
lattices with high statistics have  been obtained$^{\cite{QUENCHED}}$.
In this drastic and uncontrolled approximation the QCD calculations
are performed neglecting the contribution of the sea quarks. The only
way to estimate the quenched systematic error is to repeat the
calculations with dynamical sea quarks, but these simulations require one
or two orders of magnitude more CPU time than the quenched
ones. \par

In the last years some groups have started the simulations of full QCD
using the Hybrid Monte Carlo algorithm$^{\cite{HMCA}}$ both for
staggered and for Wilson dynamical fermions$^{\cite{UKA92}}$. In the
case of the simulations of QCD with two degenerate flavours of Wilson
fermions the mass spectrum and decay constants
calculations$^{\cite{UNQUENCHED}}$ are restricted to $16^3$ spatial
lattice size, with $\beta=5.3-5.6$ and up to $({m_{\pi} \over
m_{\rho}} )^2 \sim 0.3$. There are no results about the glueballs mass
spectrum for Wilson dynamical fermions.  \par

In these full QCD calculations the main sources of errors
have been found to be the correlations between different molecular dynamics
trajectories$ ^{\cite{GUP2}\cite{BIT92}}$ and finite size
effects$^{\cite{SIZEFIN}}$. There are preliminary studies on the correlations
for Wilson  dynamical fermions, while the finite size effects
have been studied, up to now, only for staggered fermions. \par

In order to control all possible systematic and statistical errors in
a realistic full QCD simulation,  a great amount of CPU time and a
large memory capacity are required. These are the problems that caused the
development of QCD-dedicated parallel supercomputers$^{\cite{MACHINES}}$.
The APE100 parallel machines$^{\cite{APE100}}$ promise to be a good tool to
face the problem of unquenched QCD simulations.\par

APE100 is a synchronous SIMD machine, composed of a 3-dimensional cubic
mesh of processors, with periodic boundary conditions. Each node
contains a floating point processor (MAD)$^{\cite{MAD}}$, a memory
bank of 1M 32-bit words and a network interface device. The main
advantage of this simple structure is modularity: the mesh can be
enlarged to obtain more processing power. A scalar CPU takes care of
the program flow and of all integer operations that essentially amount
to address computation and do-loop book-keeping. The MAD processor
operates on IEEE 32-bit floating point number (single precision). \par

The results presented in this paper are obtained using the single board
configuration (8 nodes disposed on a $2 \times 2 \times 2$ cube), with
a peak speed of 400 Mflops and the ``tube'' configuration (128 nodes
disposed on a $2 \times 2 \times 32$ lattice) with a peak speed of 6.4
Gflops. The next available APE100 configurations will be the 25 Gflops
``tower'' (512 nodes in a $8 \times 8 \times 8$ lattice)
and the full 100 Gflop APE100 with 2048 nodes arranged on a
$16 \times 16 \times 8$ lattice. \par

In this paper we present the implementation of the HMCA on the APE100
machines, the tests performed and the QCD results obtained up to now.
In section 2. we review the basic formalism of  unquenched QCD, then in
section 3. we present the Hybrid Monte Carlo algorithm implemented on
APE100 and we underline the parameters' tuning required and the tests
which can be performed. A short technical discussion about the single
precision aspects, the inversion algorithms and the random number
generator used will be found in section 6. A detailed analysis of the
technical aspects of the implementation of the HMCA on the APE100
machines and of the performances obtained will be presented separately$
^{\cite{TECHASP}}$. \par
In section 4. we present the tests and the results obtained using
the HMCA in the quenched approximation on $6^4$ lattices at $\beta=5.7$ and
$\beta=6.0$, while in section 5. we present the results for the full QCD case.
The latter have been obtained on $8^4$ and on $12^3 \times 32$ lattices at
$\beta=5.3$, with values of Wilson hopping parameter $k_{sea}$ lower than
that corresponding to the
strange one. Finally in section 7. we report our conclusions and
discuss the
full QCD simulations that can be faced with the different APE100
machines configurations. \par

\section{ Unquenched QCD}

The greatest obstacle to the simulation of QCD on a lattice is the inclusion of
dynamical fermions. Most of the simulations of lattice QCD have
ignored the contributions of sea quarks by using the quenched
(or ``valence'') approximation. The QCD Lagrangian is

\begin{equation}
L_{QCD} =  L_{YM} + \sum_{f=1}^{N_f} \bar \psi_f ( \Dslash + m) \psi_f \ ,
\ene
\vskip 10pt

\noindent where $L_{YM} = -{1\over 4}  F^{\mu \nu a } F^{\mu \nu a } $,
 $f$ is the flavour index and $D$ is the usual covariant
derivative. The Feynman path integral for the partition function $Z$
is:

\beg
Z =  \int {\cal D} A_{\mu} {\cal D} \bar \psi {\cal D} \psi e^{-\int d^4x
L_{QCD}} \ .
\ene
\vskip 10pt

The integration over the fermionic degrees of freedom gives

\beg
Z =  \int {\cal D} A_{\mu} [det(\Dslash + m)]^{N_f}  e^{-\int d^4x L_{YM}} \ .
\ene
\vskip 10pt
In the quenched approximation the number of flavours, $N_f$
is set to zero.
Consequently, in terms of  perturbative approach, in this
approximation one does not consider sea quark contribution to physical
observables, neglecting virtual quark loops and treating
fermions as static degrees of  freedom.  \par
It will be interesting to study the effects of the sea quarks on all
the QCD observables, and particularly on that ones that are supposed to
present strong discrepancies  with the quenched
results, such as the $\sigma$-term of the nucleon$^{\cite{QUEST},
\cite{SIGMATERM}}$, or the the $(\pi-\eta^{'})$ mass splitting.
Of course, if one could reach the physical $m_{\pi} / m_{\rho}$ value,
the most interesting observable would be the branching ratio of the
$\rho \rightarrow 2 \pi$ decay.
Up to now the numerical full QCD studies $^{\cite{UKA92}}$ are
performed at high quark masses, with statistic not comparable with the
quenched one.\par

In order to go beyond the quenched approximation, we need to calculate
the non-local fermionic determinant. A standard way to compute a
determinant requires$^{\cite{UKA88}}$ about $ N \sim O(V) $
arithmetic operations so the use of standard Monte Carlo algorithms
(like Metropolis, which requires the calculations of the fermionic
determinant for the updating of each link) is practically impossible. \par
Taking into account the effects of the sea quarks one can write
the fermionic determinant  by the introduction of a path integral over
the pseu\-do\-fermionic fields $\phi$. Let call $ M = (\Dslash + m)$,

\beg
Z =
\int {\cal D} A_{\mu} \ \biggl[det M \biggr]^{\dagger {N_f\over 2}}
\ \biggl[det M \biggr]^{ {N_f\over 2}}  \
e^{-\int d^4x L_{YM}} \
\ene
\vskip 10pt
\beg
= \int {\cal D} A_{\mu} det\biggl[M^{\dagger}M\biggr]^{{N_f\over 2}}
 \ e^{-\int d^4x L_{YM}} \
\ene
\vskip 10pt
\beg
=  \int {\cal D}A_{\mu} \ e^{-\int d^4x L_{YM}} \  \int
{\cal D}\phi {\cal D} \phi^{*} e^{-\phi^{*}[M^{\dagger}M]^{-{N_f\over 2}}
 \phi} \ .
\ene
\vskip 10pt

On the lattice the previous path integral becomes

\beg
Z =  \int {\cal D}U  \
{\cal D}\phi {\cal D} \phi^{*} \ e^{-S_G(U)- \phi[M^{\dagger}M
]^{-{N_f\over 2}} \phi^{*} }
\ene
\vskip 10pt

\noindent where $S_G(U)$ is the standard  $SU(3) $ Wilson action for
the gauge fields,

\beg
S_G(U) = {\beta \over N_{col}} \sum_{\Box} tr(I - Re(U_{\Box}))
\ene
\vskip 10pt

\noindent and the sum is over the plaquettes

\beg
U_{\Box} = U_{\mu}(x)U_{\nu}(x+\mu) U^{\dagger}_{\mu}(x+\nu)
U^{\dagger}_{\nu}(x) \ .
\ene
\vskip 10pt

One observable which is generally calculated also in full QCD
simulations is the plaquette $ \sum_{\Box} tr( Re(U_{\Box}))$
that is related to the pure gauge energy as we can see from eq.(2.8). \par

We have used the following formulation of the fermionic operator M,
that appears  in the lattice action (2.7):

\beg
M(x,y) = \delta_{x,y} - k \sum_{\mu=1}^{4} \biggl[ (1+\gamma_{\mu})
U^{\dagger}_{\mu}(x) \delta_{x, y-\mu} +
(1-\gamma_{\mu}) U_{\mu}(x-\mu) \delta_{x, y+\mu} \biggr] \ .
\ene
\vskip 10pt

In the next section we discuss the algorithm that we use to generate
full QCD configurations of link fields $\{U_{\mu}(x)\}$ distributed according
to the measure present in the path integral of eq.(2.7).
Given a set of these $\{U_{\mu}(x)\}$ configurations one can perform
a calculation of the quark propagator in order to study
the hadronic mass spectrum and the meson decay constants. \par
For example to perform a study of the meson properties we have to
compute the following correlation functions:

\beg
G_{55}(t) = \sum_{\vec{x}} < P_5 (\vec{x}, t)P_{5}^{\dagger} (\vec{0}, 0)>
\ene
\vskip 10pt

\noindent for pseudoscalar meson and

\beg
G_{kk}(t) = \sum_{k=1,3} \sum_{\vec{x}} <  V_k (\vec{x}, t)
V_{k}^{\dagger} (\vec{0}, 0) >
\ene
\vskip 10pt
\noindent for vector mesons. $P_5,\  V_k $ are the local
pseudoscalar density and vector currents respectively:

\beg
P_{5} (\vec{x}, t) = i \bar \psi (\vec{x}, t) \gamma_5 \psi (\vec{x},t)
\ene
\vskip 10pt
\beg
V_{k} (\vec{x}, t) =  \bar \psi (\vec{x}, t) \gamma_k \psi (\vec{x},t)
\ene
\vskip 10pt

\noindent where $\psi$ is  the spinor corresponding to the quark with
Wilson hopping parameter $k$, which we will call $k_{valence}$, which
can be equal or different from the $k_{sea}$ value. Usually it is
interesting to extract the hadronic mass spectrum for different value
of $k_{valence}$, at fixed $k_{sea}$ in order to study the chiral
limit. \par
In order to extract the meson masses one can fit the two point
correlation functions in the eqs.(2.11-2.12) to the expressions:

\beg
G_i(t) = {Z_i \over M_i} e^{-M_i{T\over 2}} cosh( M_i ({T\over 2} - t)),
\ \ i=55,\  kk
\ene
\vskip 10pt

\noindent where $T$ is the lattice size in time. The data for the
mesonic correlation functions can be fitted to the previous
functions, obtaining $M_i$. The time interval which should be used
in the fit can be chosen by checking that the effective mass
$ m_{eff}(t) = log(G(t)/G(t+1)) $ reaches a plateau value in the time
interval considered which agrees with the result of the fit.
Preliminary results for the mesons mass spectrum are reported in
section 5. \par
In the next section we present the Hybrid Monte Carlo algorithm
which is used to obtain full QCD configurations.\par

\section{ The Hybrid Monte Carlo Algorithm. }

In this section we describe the HMCA that we are using for
the full QCD simulation. The HMCA generates gauge
configurations and pseudofermionic fields distributed according to the
measure given in eq.(2.7). This is done at first suggesting a change
to the whole lattice using the $\Phi$ algorithm$^{\cite{HMCA}}$,
 and then globally
accepting or rejecting this change in the Metropolis step.
These two steps  are cyclically repeated.\par
In the first step, in order to obtain the proposal for the new
configuration in the $\Phi$ algorithm, one introduces an Hamiltonian
${ \cal H}(P,U)$  which depends on the link variables $P_{\mu}(x)$ and
$U_{\mu}(x)$:

\beg
{ \cal H }(P,U) = {1\over 2} tr (P^2 ) + S_G(U) +
\phi^{\dagger}(M^{\dagger}M)^{-1}\phi
\ene
\vskip 10pt

\noindent $U_{\mu}(x)$ are the standard gauge link variables, while
$P_{\mu}(x)$ play the role of Hamiltonian  conjugate momenta.  The sum of the
second and the third term in equation (3.1) is exactly the action that appears
in the exponential of eq.(2.7).  The physical observables are not
affected by the new term depending on $P$, because it can be factored out
becoming gaussian integral that only affects the normalization. \par
The Hamiltonian variables are initialized in the following way: to generate
the pseudofermionic $\phi$ fields we extract random complex numbers $\eta$
distributed as $P(\eta)\propto e^{(-\eta^{\dagger}\eta)}$. In this way
$\phi=M^{\dagger}\eta$ is distributed according to the pseudofermionic
measure of the partition
function (2.7). The $\phi$ fields have no dynamic and thus there are no
$\phi$-conjugate momenta into the Hamiltonian.\par
In order to generate the $P_{\mu}(x)$ variables we extract real random numbers
$g^a_{\mu}(x)$ which are drawn with probability distribution
$P(g) \propto e^{(-g^{2})} $, and then we construct:

\beg P_{\mu}(x) =
\sum_{a=1}^{8} g^a_{\mu}(x) \lambda_a
\ene
\vskip 10pt
\noindent where $\lambda^a$ are the Gell-Mann matrices. We want to note that
both $P_{\mu}(x)$ and $\phi$ variables are generated randomly to ensure
ergodicity.\par

The $(P,U)$ configuration is evolved according to the Hamiltonian equations of
motion, which in the QCD case  can be obtained in the following way.
In order for the matrix $U_{\mu}(x)$ to remain an element of $SU(3)$,
if  $P_{\mu}(x)$ momenta are  traceless and hermitian operators, the
equation of motion for $U_{\mu}(x)$ must be
\begin{equation}
 \dot{U}_{\mu}(x) =  i P_{\mu}(x) U_{\mu}(x)  \ .
\end{equation}
To obtain the evolution equation for the $P_{\mu}(x)$, we require
the Hamiltonian to be a constant of the motion$^{\cite{HMCA}}$,
$\dot{{\cal H}} = 0$. In this way we find
\begin{eqnarray}
 \dot{P}_{\mu}(x) =   - {\delta S(U) \over \delta U} + \phi^{*}
(M^{\dagger}M)^{-1} \left[ M^{\dagger} {\delta M \over \delta U}
+ {\delta M^{\dagger} \over \delta U} M \right] (M^{\dagger}M)^{-1} \phi \ .
\nonumber
\\
&&
\end{eqnarray}
\vskip 10pt
To obtain a ``trajectory'' we have to refresh the $P_{\mu}(x)$ and the
$\phi(x)$ variables and then to evolve the $(U,P)$ configuration for a fixed
time $\tau$. In the numerical integration of the Hamiltonian equations we
introduce a finite time step $dt$. The number of molecular dynamics steps is
$N_{MD}$, and the trajectory length is $\tau=dt \times N_{MD}$.\par
We have initially tested the implementation of this algorithm in the absence of
fermionic fields. Under these conditions the algorithm is used to produce
quenched pure gauge configurations. These configurations can be statistically
compared with those ones obtained with well tested existing Metropolis program.
\par
 Defining the staples

\begin{eqnarray}
 V_{\mu} (x) = && \sum_{\nu} \Bigl (
 U_{\nu}(x+\mu,t) U^{\dagger}_{\mu}(x+\nu,t)U^{\dagger}_{\nu}(x,t)
 \nonumber
  \\
 && +  U^{\dagger}_{\nu}(x+\mu-\nu,t)
 U^{\dagger}_{\mu}(x-\nu,t) U_{\nu}(x-\nu,t) \Bigr)  \ ,
\end{eqnarray}
\vskip 10pt

\noindent the discrete evolution equations for the ``quenched'' case can
 be written as:

\begin{eqnarray}
& U_{\mu}(x,t+dt) =  e^{i P_{\mu}(x,t)dt} U_{\mu}(x,t)
\\
& P_{\mu}(x,t+dt) = P_{\mu}(x,t) + {i \beta \over N} dt \hat{T} \left[
U_{\mu}(x,t) V_{\mu}(x) \right] \ ,
\end{eqnarray}
\vskip 10pt

\noindent where $\hat{T}$ is the traceless antihermitian projector:

\beg
\hat{T} (A) = (A-A^{\dagger}) - {1 \over N} { \bf I } \ Tr (A-A^{\dagger}) \ .
\ene
\vskip 10pt

The discrete evolutions equations used in the full case are

\beg
 U_{\mu}(x,t+dt) =  e^{i P_{\mu}(x,t)dt} U_{\mu}(x,t)
\ene
$$
P_{\mu}(x,t+dt) = P_{\mu}(x,t)
$$
\beg
 + i  dt T  \left[
{ \beta \over 6} U_{\mu}(x,t) V_{\mu}(x)
 - k_{sea} \cdot tr_{Dirac} ( U_{\mu}(x,t) F_{\mu}(x)) \right]
\ene
\vskip 10pt

\noindent where in the third term in eq.(2.21) the trace is over the Dirac
indices and

\beg
F_{\mu}(x) =  (1+\gamma_{\mu}) Y(x+\mu) \chi^{\dagger} (x) +
(1-\gamma_{\mu}) \chi(x+\mu) Y^{\dagger}(x)
\ene
with
\begin{eqnarray}
& \chi(x) = & \left( M^{\dagger} M \right)^{-1} \phi
\\
& Y(x) = & M \left( M^{\dagger} M \right)^{-1} \phi \ .
\end{eqnarray}

\vskip 10pt
\vskip 10pt
\vskip 10pt
In the previous equations we used an expansion of the exponential up
to fourth order, which guarantees the time reversibility\footnote{
We have also tried sixth order in the exponential expansion, without any
significative difference in reversibility.}. At each step
of the molecular dynamics evolution we normalized the $SU(3)$ link matrices
 $U_{\mu}$.

In the time integration of the Hamiltonian equation we use a leap-frog
algorithm $^{\cite{HMCA},\cite{KEN88}}$, which is time reversible and
area preserving in the phase space. The reversibility assures the
validity of the principle of detailed balance.
The leap-frog algorithm disentangles the update of $U$ and $P$ variables,
that are no more evaluated at the same time: we perform an initial halfstep
of size $dt/2$, and then a chain of $N_{MD} -1$ steps of size $dt$,
alternatively in $U$ and $P$.
Eventually, a last halfstep for the $P$ momenta produces the $U(\tau),P(\tau)$
configuration.
The deterministically proposed configuration over the whole lattice
$(U^{\prime},P^{\prime})$ is then globally accepted or rejected with
a probability

\beg
 \varrho = \min \left( 1,e^{ - ({ \cal H}(P^{\prime},U^{\prime}) -
                                                { \cal H}(P,U))} \right)
 \ .
\ene
\vskip 10pt

If we consider the usual Metropolis algorithm applied to full QCD we
see that in a single sweep we need to make as many inversions of the
fermionic operator as the number of the links we want to update, ($ {
\cal O } $(V)), while with a trajectory of the HMCA we obtain the
probability distribution (2.7) with only $ { \cal O } $(1) inversions.
The main difference between the two algorithms is that in the
Metropolis case the links of the new configuration are chosen randomly
and the accept/reject condition is imposed link by link (locally),
while in the HMCA the new configuration is obtained by the previous
deterministic equations (3.3,3.4) and the accept/reject condition is
imposed over the whole configuration (globally).
\vskip 10pt
A useful test for the implementation of the HMCA is the
identity $^{\cite{CREUTZ}}$:
\beg
< e^{- \delta {\cal H}} > = 1 \ .
\ene
\vskip 10pt
where $\delta H = H(P^{\prime},U^{\prime})-H(P,U) = H^{\prime}-H$.
This identity follows from the validity of the Liouville theorem
of the Hamiltonian evolution $ {\cal D} U {\cal D} P =
{\cal D} U^{'} {\cal D} P^{'} $ and becomes exact when the system
has reached the equilibrium:

\beg
\label{creutz_ave}
1 = {Z \over Z} = {1\over Z} \int {\cal D} U^{'} {\cal D} P^{'} e^{-{\cal
H}^{'}}
= {1\over Z} \int {\cal D} U {\cal D} P e^{-{\cal H}}  e^{{\cal H}-{\cal
H}^{'}} \ .
\ene
\vskip 10pt
The (\ref{creutz_ave}) is satisfied for every choice of the algorithm
parameters and permits a fine control on the proposed configuration.
This identity is not satisfied if the iterative inversion algorithm
for the pseudofermionic operator $M^{\dagger} M$ is not successfully completed,
with a strong enough degree of convergence. Then we can check the validity of
eq.(3.15) to test the thermalization of our configuration and to fix the
value of the residue of the inversion. In the previous identity the
average is taken over all the trajectories (accepted or not accepted).
A consequence of equation (3.15)
is that we expect$^{\cite{CREUTZ}}$ that  \par
\beg
< {\cal H}^{'} - {\cal H} > \ \ge \ 0 \ .
\ene
\vskip 10pt

We use periodic boundary condition on the gauge fields. On the
fermionic field we use periodic boundary condition in the time
direction and periodic or antiperiodic spatial boundary
condition. \par

An important feature of the HMCA is the parameter's tuning.
There are many parameters that have to be tuned: the free technical
parameter are $N_{MD}$, the number of steps of a molecular dynamics
(MD) trajectory and $dt$, the length of each step. The free physical ones
 are $\beta = {6 \over g^2}$ and the hopping parameter
$k_{sea}$ into the fermionic operator $M$ related to the sea quark
mass. The physical parameters appear both in the evolution equations
(3.9-3.10) and in the accept/reject Hamiltonian of eq.(3.14), so a further
possible tuning, which has not been  explored in the present paper, is the
usage$^{\cite{HMCA}}$ in the evolution equations of a $\beta_{MD}$ and
a $k_{MD}$ different from the ones used in the accept/reject
Hamiltonian. \par
In our simulations we have fixed $\beta_{MD} = \beta$, $k_{MD} = k_{sea}$ and
the value of $\tau = dt \times N_{MD}$, and we have studied the behaviour
of the  acceptance as a function of $dt$.
We check that in an exact algorithm like the HMCA  the value of
the plaquette is independent from $dt$. \par

In the full HMCA evolution equations (3.9,3.10), we need the
quantities reported in eqs.(3.12,3.13). To obtain $\chi(x)$ and then
$Y(x)$ we have to perform the inversion of the fermionic operator
$M^{\dagger} M$.

\beg
M^{\dagger} M \chi = \phi \ .
\ene
\vskip 10pt

This is the most time-consuming part of the algorithm
and must be repeated $N_{MD}$ times for each trajectory.
To solve eq.(3.18) we use the Conjugate Gradient algorithm
described in section 6. We stop the iterative algorithm when

\beg
\frac{ ((M^{\dagger} M \chi - \phi), (M^{\dagger} M \chi - \phi))}
{ (\chi,\chi) } \ \le \ R
\ene
\vskip 10pt

\noindent is lower than a fixed R value$^{\cite{GUPTA1}}$.

\section { Quenched Results }

In this section we summarize our results obtained using both
the Hybrid Monte Carlo and Hybrid Molecular Dynamics algorithm$^{\cite{HMCA}}$
in the pure gauge case. We call Hybrid Molecular Dynamics algorithm
an HMCA without the acceptance step. While the HMCA is
an exact algorithm, the observables measured on configurations
produced with HMD depend on the step size $dt$ of the trajectory. \par
In order to test our HMCA we performed numerical simulations
with only the pure gauge part of the action. This means that the
proposed configuration is obtained using the Hamiltonian equations
without pseudofermions degrees of freedom, see eqs.(3.6-3.7). \par

The simplest test we performed is the following: we check that the HMC
pure gauge algorithm reproduces the same value of the plaquette as the
Metropolis algorithm in a quenched simulation. Then we analyze the
behaviour of the acceptance as a function of the step size $dt$ and
finally we verify that the identity (3.15) is satisfied for all our HMC
simulations. \par

We perform numerical simulation on a $6^4$ lattice at two different
value of $\beta$, $\beta=5.7$ and $\beta=6.0$. For each value of
$\beta$ we studied the HMCA behaviour changing the free
parameter $dt$. We choose the number of steps $N_{MD}$ in order to
maintain a fixed trajectory length, $\tau = dt \times N_{MD} \sim 0.5$.
The results for the acceptance rates, for the plaquette values
and for $<e^{-\delta {\cal  H}} >$ are reported in Table 1. \par

We remark that the value of the plaquette is compa\-ti\-ble with the
quen\-ched
Metropolis result: for example, on a $6^4$ lattice, from 4000 sweeps
of the Metropolis algorithm after 1000 sweeps of thermalization we
obtain $P(\beta=6.0) = 0.5946(4) $, in agreement with the HMCA pure
gauge result $ P(\beta=6.0) = 0.5951(2) $. Analogously at $\beta=5.7$
the Metropolis data is $P(\beta=5.7) = 0.5484(7)$ and the HMCA result
is 0.5487(6). \par
We note, as expected for an exact algorithm like HMCA, that within
the errors the value of the plaquette does not depend on
$dt$. \par
The errors quoted on the plaquette values have been obtained using the
jacknife method: decreasing the number of the bins we analyze the
behaviour of the error and we take the plateau value. In figure 1 we
report the $\beta=6.0$ plaquette as a function of the number of
trajectories. We observe the presence of correlations on different
scales. This is a well note problem of the HMCA and looking at this
figure we can conclude that our errors may be underestimated. For this
reason usually people prefer to use the Metropolis algorithm rather than
HMCA to produce quenched uncorrelated configurations.\par

Looking at the data of Table 1 we note a sharp variation of the
acceptance as a function of $dt$. An increase in the acceptance needs
a decrement of the value of $dt$. In order to maintain a given acceptance
the step size $ dt $ must be reduced when going to larger
lattices, as we have found in some test simulations performed on $18^3
\times 32$ lattices$^{\cite{CREUTZ}\cite{GUPTU}}$.

\begin{table}
\begin{tabular}{||c|c|c|c|c|c|c||} \hline
$\beta$   &   $dt$   &   $N_{MD}$  & $N_{traj}$ & Accep. & Plaq. &
$<e^{- \delta \cal H}>$
\\ \hline\hline
$5.7$ & $0.075$ & $6$  & $16000$ & $16.5\% $ & $0.552(2)$ & 1.2(3)
\\ \hline
$5.7$ & $0.050$ & $10$ & $2000$  & $56.7\%$  & $0.5491(6)$ & 1.07(6)
\\ \hline
$5.7$ & $0.025$ & $20$ & $4000$  & $89.5\%$  & $0.5487(6)$  & 1.002(3)
\\ \hline
\hline
\\ \hline
$6.0$ & $0.075$ & $6$  & $8000$  & $13.0\%$ & $0.5950(3)$ & 0.7(1)
\\ \hline
$6.0$ & $0.050$ & $10$ & $2000$  & $54.4\%$ & $0.5947(2)$ & 0.99(3)
\\ \hline
$6.0$ & $0.025$ & $20$ & $2000$  & $87.4\%$ & $0.5951(2)$ & 1.000(6)
\\ \hline
\end{tabular}
\caption{
Hybrid Monte Carlo pure gauge simulations. All the simulations have
been performed on a $6^4$ lattice.}
\protect\label{SIMULATIO}
\end{table}

\par

\begin{table}
\begin{tabular}{||c|c|c|c|c||} \hline
$\beta$   &   $dt$   &   $N_{MD}$  & $N_{traj}$ & Plaq. \\ \hline\hline
$5.7$  &   $0.100$  &   $5$        & $1000$     & $0.5315(15) $ \\ \hline
$5.7$  &   $0.075$  &   $6$        & $1000$     & $0.5383(13) $ \\ \hline
$5.7$  &   $0.050$  &   $10$       & $1000$     & $0.5438(11) $ \\ \hline
$5.7$  &   $0.025$  &   $20$       & $1000$     & $0.5500(10) $ \\ \hline
\hline
$6.0$  &   $0.100$  &   $5$        & $1000$     & $0.5803(5)$ \\ \hline
$6.0$  &   $0.075$  &   $6$        & $1000$     & $0.5865(5)$ \\ \hline
$6.0$  &   $0.050$  &   $10$       & $1000$     & $0.5917(4)$ \\ \hline
$6.0$  &   $0.025$  &   $20$       & $1000$     & $0.5945(5)$ \\ \hline
\end{tabular}
\caption{
Hybrid Molecular Dynamics pure gauge simulations. All the simulations
have been performed on a $6^4$ lattice.
}
\protect\label{HYBRID}
\end{table}

\par

As we underline in section 3, we checked, following ref.\cite{CREUTZ},
that our HMCA simulations satisfy the identity (3.15).
In this way we are sure that the thermalization has been done correctly
and all the properties of the Hamiltonian equations are
conserved in the implementation. In the pure gauge simulations
we obtain the good results reported in the seventh column of Table 1. \par
We also perform simulations using the HMDA, on a $6^4$ lattice at
$\beta=5.7$ and $\beta=6.0$. The results are summarized in Table 2.
{}From the data of Table 2 we note, as expected, that the value of the
plaquette obtained with a HMD algorithm depends on $dt$. In fact the
plaquette approaches  from lower values the quenched result obtained
using an HMCA, as noted in  ref.\cite{GUPTU}. \par
\vskip 10pt
\vskip 10pt
\vskip 10pt

\section { Unquenched Results }
In this section we present our tests and first results obtained using the
APE100 machine for full QCD simulation.\par
All the simulations have been performed at $\beta=5.3$ and $k_{sea} =
0.158$ on $8^4$ and on $12^3 \times 32$ lattices. On $8^4$ we use
different values of $dt$ and $N_{MD}$. In all the simulations we use
periodic boundary conditions on the fermionic degrees of freedom,
while for $dt=0.02, \ N_{MD}=25$ we use both periodic and antiperiodic
spatial boundary conditions on the fermions. \par

Our tests consist in the calculation of the average value of the three
different contributions at the Hamiltonian. We check that the
$P_{\mu}$ link variables initialized according to eq.(3.2) at the begin
of each trajectory, are distributed according to eq.(3.2) also at the end
of each trajectories and that $< {1\over 2} {1\over N_{link}}
\sum_{x,\mu} tr(P^2_{\mu}(x))> = 4.0$. We check that the pseudofermionic
contribution to the action has an average equals to 12.0 and that the
plaquette values are statistically comparable with the ones presented in the
literature$^{\cite{GUPTA1}\cite{GUP2}}$.  \par

\begin{table}
\begin{tabular}{||c|c|c|c|c|c|c|c||} \hline
$dt$ & $N_{MD}$ & Ac. & $<\delta \cal H>$ & $<e^{- \delta \cal H}>$
&  $< \phi^{\dagger} (M^{\dagger} M)^{-1} \phi >$ &
$<{1\over 2} tr(P^2)>$ & Plaq.\\ \hline\hline
$0.04$ & $13$ & $54\%$  & $0.81(5)$ & 1.03(6) & 12.005(2) & 3.995(1)
& 0.489(1)  \\ \hline
$0.02$ & $25$ & $88\% $ & $0.06(1)$ & 1.00(2) & 12.001(2) & 3.999(1)
& 0.4894(8)  \\ \hline
$0.01$ & $50$ & $98\%$  & $0.003(4)$ & 1.001(4) & 12.002(2) & 3.9991(7)
& 0.489(1)
\\ \hline
$0.02$ & $25$ & $89\% $ & $0.05(1)$ & 1.01(1) & 12.002(2) & 3.9998(5)
& 0.490(1) \\ \hline
\hline
\end{tabular}
\caption{
Hybrid Monte Carlo full QCD simulations on $8^4$ lattice, $\beta=5.3$,
$k_{sea}=0.158$.
The data in the first three rows are obtained using periodic boundary
conditions on the fermionic degrees of freedom, while the last row
contains data obtained using spatial antiperiodic and temporal
periodic boundary conditions on the fermions.
The number of trajectories for these simulations are 890, 678, 640, 967
from the top to the bottom of the Table respectively.
}
\protect\label{UNQUE1}
\end{table}

To invert the pseudofermionic operator $M^{\dagger} M$ we use a
Conjugate Gradient algorithm (see section 6.). We decrease the value
of R (see eq.(3.19)) to obtain a variation in $\delta {\cal H}$  of
the order of one percent. In the simulation reported in this paper we obtain
this result for $R \le 10^{-13}$. For example for a given $12^3 \times 32$
configuration (of the set used for the results in Table 4),varying the value
of R in the stop condition of the CG algorithm
we have found the $\delta H$ results reported in Table 5. We note
that $1\%$ accuracy in $\delta {\cal H}$ requires $R
 \le 10^{-13}$. \par
In these initial test we start our run from random $U_{\mu}$ configuration.
Indeed starting from a configuration with a critical value of the hopping
parameter $k_c^{init}$ we want to
reach the final thermalized configuration with $k_c = k_c(\beta)$.
If in the thermalization we don't want to pass through a
 configuration with $k_c$ equals to
the value of $k_{sea}$ of our unquenched simulation we have to
start from a configuration warmer than the one we want to reach after
the thermalization. This means that we can start our full QCD run
also from a configuration just a little warmer than the final one,
for example a pure gauge configuration with these properties.
Indeed following the studies of ref.\cite{GUPTA1},  for a given unquenched
configuration we can estimate the equivalent gauge coupling for the quenched
theory which has the same lattice scale. \par

The second and the fourth rows of Table 3 contain the data obtained
using periodic and antiperiodic spatial boundary condition
respectively. In this case the results are statistically comparable
but if we look at figure 2. the behaviour of the plaquette versus
the number of trajectories seems to be different. \par

First results for the mass spectrum on $12^3 \times 32$ lattice
are obtained at $\beta=5.3$. The value of the
HMCA parameter are those
reported in Table 4. In this case we have 1180 preliminary
trajectories. APE100 is currently running $^{\cite{NEWRUN}}$
at higher value of $k_{sea}$. \par
In figure 3. we report the behaviour of the plaquette,
of $ \phi^{\dagger} (M^{\dagger} M)^{-1} \phi $, of $-\delta {\cal H}$
and of $ {1\over 2} {1\over N_{link}} \sum_{x,\mu} tr(P^2_{\mu}(x)) $
for the $12^3 \times 32$ simulation as functions of the
trajectory number. As can be seen from table 4 the results
for the testing variables are in
good agreement with theoretical predictions and with the plaquette
results reported in literature$^{\cite{GUPTA1}}$ .
In figure 4 we report the average of the correlation functions for
the pseudoscalar and for the vector mesons at $k_{valence} =
 k_{sea} = 0.158$.
In figure 5 we report the behaviour of the effective mass for these
particles. \par
In figure 6 we report the time history of pseudoscalar and
vectorial correlation at
euclidean time separation 11.
{}From this figure
we observe the presence of autocorrelation on different scales$^{\cite{GUP2}
\cite{BIT92}}$.\par
We start the run with 30 HMD trajectories, then we perform 180 thermalization
HMC trajectories. At this point we invert the fermionic operator to obtain
a propagator for a given $k_{valence}$ every 5 trajectories.\par
In this paper we report a preliminary analysis of the pseudoscalar and
vectorial
correlation functions (see eq (2.11-2.12)) obtained on this ensemble of
 unquenched configurations (see figure 4, 5, 6).\par

\vskip 10pt

\begin{table}
\begin{tabular}{||c|c|c|c|c|c|c|c||} \hline
$dt$ & $N_{MD}$ & Ac. & $<\delta \cal H>$ & $<e^{- \delta \cal H}>$
&  $< \phi^{\dagger} (M^{\dagger} M)^{-1} \phi >$ &
$<{1\over 2} tr(P^2)>$ & Plaq.\\ \hline\hline
$0.01$ & $60$ & $92\%$ & $0.04(1)$ & 1.01(1) &  12.0003(4) & 4.0001(2)
& 0.4904(2)
\\ \hline \hline
\end{tabular}
\caption{
Hybrid Monte Carlo full QCD simulations on $12^3 \times 32$ lattice,
$\beta=5.3$, $k_{sea}=0.158$, with periodic boundary conditions on the
fermionic degrees of freedom.
}
\protect\label{UNQUE2}
\end{table}

\par

\begin{table}
\begin{tabular}{||c|c||} \hline
$R$ &  $\delta \cal H$ \\ \hline\hline
1.0e-06  &     -0.8375   \\ \hline
1.0e-07  &     -0.6997   \\ \hline
1.0e-08  &      0.1123   \\ \hline
1.0e-09  &      0.1842   \\ \hline
1.0e-10  &      0.1878   \\ \hline
1.0e-12  &      0.1712   \\ \hline
5.0e-13  &      0.1747   \\ \hline
1.0e-13  &      0.1763   \\ \hline
7.0e-14  &      0.1767   \\ \hline
\hline
\end{tabular}
\caption{ The value of $\delta {\cal H}$ as a function of the precision
R (see equation 3.19) of the inversion for a fixed configuration in
the set described in Table 4.
}
\protect\label{RDH}
\end{table}

\section{Technical Aspects.}

In the implementation of the algorithm on the APE100 machine, some
care must be devoted to possible problems due to the single-precision
of the floating point units. With the lattices used in the actual full
QCD simulations, this type of problems are not so
 important in iterative algorithms with stable solution,
 where, if convergence occurs, the fixed point
solution is reached anyway, eventually with a larger number of iterations.
The difference between single and double precision has to be paied
 in CPU time. However some improvements can be obtained taking care
in the implementation of the inversion algorithm.
\par
To solve the equation

\begin{equation}
 { \bf M^{\dagger} M } \chi = \phi
\end{equation}
defining the residue at the $n$ iterative step
\begin{equation}
R_n =  { \bf M^{\dagger} M } \chi_n -  \phi
\end{equation}
the Conjugate Gradient (CG) iterative equations are:

\begin{equation}
\chi_0 = 0 \ , \ \ \  P_0 = R_0
\ene
\beg
a_n = {(R_n,R_n) \over (M P_n, M P_n) }
\ene
\beg
\chi_{n+1} = \chi_n + a_n P_n
\ene
\beg
R_{n+1} = R_n - a_n (M^{\dagger}M) \ P_n
\ene
\beg
b_n = {(R_{n+1},R_{n+1}) \over (R_n, R_n) }
\ene
\beg
P_{n+1} = R_{n+1} + b_n P_n
\ene

\noindent where  $ \chi_n $ is the solution after $n$ steps.

To compute the CG equations (6.4) and
(6.7), we need to perform a global sum.
On APE100 we use a tree algorithm.
While a standard algorithm adds the items sequentially in a
variable, our tree algorithm consists in summing the items in pair and
in iterating this operation until the total sum is reached.
Thus the tree algorithm is designed to sum only numbers of the same magnitude
reducing the rounding errors.\par
We are also checking on APE100 machines a new algorithm proposed by Valerio
Parisi$^{\cite{VP}}$ to correct for rounding errors in summation.
This iterative algorithm uses a variable to dynamically backing up the
informations lost in the sum.\par
More details of these algorithms will be presented in ref.\cite{TECHASP}. \par

While in the inversion algorithm the rounding errors can reduce the
convergence velocity, in the calculations of $\delta { \cal H}$ they can
cause systematic errors, because $\delta { \cal H}$ is used in the
accept/reject condition.\par

Another technical aspect that we faced is the random number generation.
In the HMCA simulation we need to initialize the $ P_{ \mu } $ and the
$ \phi $ variables at the start of each new trajectory.
With large lattice for a single thermalized configuration we have
to generate  $ { \cal O } (10^{10}) $ random numbers.
Although on the APE100 machines a random number generator is available
and used in the quenched simulations since several years,
we have been lead to create a new generator of real random number with
different period, correlation and speed properties, in order to check
the possible effects on the physical results.\par

The problem with APE100 is that the machine only works
with 32 bits real numbers and allows a limited amount of mixed
operations between real and integer numbers, while  most of the tested
random generators use integers in the algorithm and then convert the
result into real numbers.
We have designed\footnote{We thank G. Parisi for all the long
and useful discussion we had with him about this subject.}
a new generator$^{\cite{RANGEN}}$ combining three different algorithms,
taking care that
the rounding error effects do not ruin the statistical and long period
properties. The mixing of the three generators can produce numbers
that are more independent than those produced by each algorithm.
In order to test the implementation of the evolution equations (3.9,
3.10) and the good properties of the new random number generator we
have checked that, during the whole evolution, the $ P_{ \mu }$ variables
follow a gaussian distribution. \par

\section{Full QCD on APE100 Machines.}
As we said in the Introduction,
the APE100 machines are parallel
supercomputers with modular building
blocks. At present we are using the ``tube'' configuration, which is
composed by 128 processors (nodes, arranged on a $2 \times 2 \times
32$ three-dimensional structure) and has a peak speed of 6.4 Gflop. In
the next future the ``tower'' configuration will be available, which
is composed of 512 nodes, arranged in a $8 \times 8 \times 8$
configuration with a peak speed of 25.6 Gflop. Each node has a memory
capacity of 4 Mbytes ( $10^6$ real 32-bit words). With this amount of
memory the maximum lattice size we can use in the case of a full QCD
simulation on the ``tube'' is  $ 20^3 \times 32 $ while on the
``tower'' is  $ 32^4$. \par

The code as it is implemented can achieve a performance of 60 \% of
the peak speed on the most time consuming part (the inversion of the
pseudofermionic $ M^{ \dagger } M $ operator and the staple
calculations). We are implementing a more efficient code using the
features of the next APE100 compiler.\par

In the $12^3 \times 32$ simulations reported in Section 5.
we obtain an high acceptance ($92\%$). From the studies on the
$8^4$ lattice we see that a good value of the acceptance can be reach
going from $dt=0.01, N_{MD}=60$ to $dt=0.03, N_{MD}=20$. In this way
we gain a factor three in the CPU time and
we expect to obtain with a ``tube'' something like 1000 full QCD
trajectories in 8 days on a $12^3 \times 32$ lattice with a quark mass in the
strange quark region. Taking into account the
autocorrelation modes we can generate $5\div 10$ independent
configurations in 8 days.  \par

Looking at all the results of full QCD spectroscopy$^{\cite{UKA92}}$,
we note that the more realistic simulations with two flavour Wilson
fermions have been performed  on  $16^3 \times 32 $ lattices with low
statistics (see for example Table 1 of ref. \cite{UKA92}).
Looking for example at the results of ref.\cite{BIT92}, we note the
large correlations that affect the time history of the pion propagator
of fig.2. This is the same situation of fig.2.a of ref.\cite{GUP2}.

This means that, more long runs have to be performed to
deeply understanding the long scale correlations  before  obtaining realistic
results. This is the first goal that we want to reach.\par

With a ``tube'' we are running  on  a  $ 12^3
\times 32 $ lattice at $\beta=5.3$ and we are performing measures
on the hadronic and glueballs mass spectrum.
 With a ``tower'' or with the full 100 Gflop, 2048 nodes machine,
we can simulate
full QCD on large lattices, comparable with those used today for
staggered fermions. Large lattices are needed to control
the errors coming from  finite size effects. The study of this systematic
error in the full QCD simulations will be another point to be
understood. \par

With these powerful machines will be possible to repeat the calculations
with dynamical see quarks of all the physical quantities (mass spectrum, decay
constant, matrix elements, etc.) obtaining eventually the correct full QCD
results.\par
\vskip 10pt
\vskip 10pt

\centerline {\bf{ Acknowledgements  }}
We thank the APE100 Collaboration that constantly have supported our activity.
We acknowledge interesting discussion with E.~Laermann, T.~Lippert,
C.~Ungarelli, T.~Vladikas, D.~Weingarten.
We warmly thanks N.~Cabib\-bo, G.~Martinelli and G.~Parisi for many
 suggestions.

\par
\vfill
\newpage
{\bf FIGURE CAPTIONS}\par

\begin {itemize}
\item
[Figure 1] The plaquette in a quenched (HMC) simulation as a function of
the number of trajectories on a $6^4$ lattice for the simulation at
$\beta =6.0$, $dt=0.025$, $N_{MD} =20$.
\item
[Figure 2.a] The plaquette in a full (HMC) simulation with periodic
boundary conditions as a function of the number of
trajectories on a $8^4$ lattice for the simulation at  $\beta =5.3$,
$dt=0.02$, $N_{MD} =25$ $k_{sea} =0.158$.
\item
[Figure 2.b] Same as fig. 2a for
anti periodic spatial
boundary  conditions as a function of the number of
trajectories on a $8^4$ lattice for the simulation at  $\beta =5.3$,
$dt=0.02$, $N_{MD} =25$ $k_{sea} =0.158$.
\item
[Figure 3] The plaquette (a), the
$ \phi^{\dagger} (M^{\dagger} M)^{-1} \phi $ (b), the
 $-\delta {\cal H}$ (c), and the $\ \ \ \ \ \ \ $
$ {1\over 2} {1\over N_{link}} \sum_{x,mu}
tr(P^2_{\mu}(x)) $ (d) in a full (HMC) simulation with periodic
boundary condition as a function of the number of
trajectories on a $12^3\times 32$ lattice for the simulation at
$\beta =5.3$,
$dt=0.01$, $N_{MD} =60$ $k_{sea} =0.158$.
\item
[Figure 4] The $G_{55}(t)$ correlation function of eq.(2.11) (a),and  the
$G_{kk}(t)$ correlation function of eq.(2.12) (b)
in a full (HMC) simulation with periodic boundary conditions
on a $12^3\times 32$ lattice for the simulation at  $\beta =5.3$,
$dt=0.01$, $N_{MD} =60$ $k_{sea} = 0.158$.
\item
[Figure 5] The pseudoscalar effective mass (a), and vectorial effective
mass (b)
in a full (HMC) simulation with periodic
boundary conditions  on a $12^3\times 32$ lattice for the
simulation at  $\beta =5.3$,
$dt=0.01$, $N_{MD} =60$ $k_{sea} =0.158$.
\item
[Figure 6] The $G_{55}(t)$ correlation function  (a) and the
 $G_{kk}(t)$ correlation function (b) as a function of trajectories number
for euclidean time separation 11
in a full (HMC) simulation with periodic
boundary conditions  on a $12^3\times 32$ lattice for the
simulation at  $\beta =5.3$,
$dt=0.01$, $N_{MD} =60$ $k_{sea} =0.158$.
\end {itemize}

\end{document}